\documentstyle[prl,aps]{revtex}
\begin{document}
\draft
\twocolumn[\hsize\textwidth\columnwidth\hsize\csname @twocolumnfalse\endcsname
\title{Drude weight and total optical weight in a 
       $\bbox{t}$-$\bbox{t}^{\bbox{\prime}}$-$\bbox{J}$ model}
\author{Gregory C. Psaltakis}
\address{Department of Physics, University of Crete, and Research Center 
         of Crete, Heraklion, GR-71003, Greece}
\maketitle
\begin{abstract}
We study the Drude weight $D$ and the total optical weight $K$ for a 
$t$-$t^{\prime}$-$J$ model on a square lattice that exhibits a metallic
phase-modulated antiferromagnetic ground state close to half-filling. Within a 
suitable $1/N$ expansion that includes leading quantum-fluctuation effects, 
$D$ and $K$ are found to increase linearly with small hole doping away 
from the Mott metal-insulator transition point at half-filling. The slow 
zero-sound velocity near the latter transition identifies with the velocity of 
the lower-energy branch of the twofold excitation spectrum. At higher
doping values, $D$ and $K$ eventually saturate and then start to decrease. 
These features are in qualitative agreement with optical conductivity 
measurements in doped antiferromagnets.
\end{abstract}
\vskip2pc]

\narrowtext
\section{Introduction}
\label{sec:intro}

The optical properties of the metallic state that emerges upon doping the 
insulating Heisenberg antiferromagnet with mobile holes are of great interest 
because such a system, currently described by a $t$-$t^{\prime}$-$J$ model, is 
believed to capture low-energy physics of the high-$T_{c}$ superconducting 
copper-oxide layers \cite{review94}. In the context of the latter model we 
have recently calculated \cite{Psaltakis95}, using a suitable $1/N$ expansion, 
the real part of the finite-frequency optical conductivity, i.e., the optical 
absorption $\sigma(\omega)$, and found that its magnitude scales with small 
hole concentration while its low-frequency line shape displays a structureless 
broadband regime. These results are in qualitative agreement with the generic 
properties of the midinfrared band observed in optical conductivity 
measurements \cite{Cooper90,Orenstein90,Uchida91} in doped antiferromagnets. 
Here we extend our earlier work \cite{Psaltakis95} by presenting results 
for the  Drude weight $D$ and the total optical weight $K$ that determine the 
zero-frequency response of the system and the optical conductivity sum rule, 
respectively. As it is well-known \cite{Kohn64}, the Drude weight $D$ 
serves as an order parameter for the Mott metal-insulator transition occurring 
in such a system, hence the significance of its dependence on hole 
concentration and relevant coupling constants. Furthermore, the fraction of the
total optical weight residing at zero frequency, i.e., the fraction $D/K$, 
defines the inverse of the mass enhancement factor. The latter quantity is a 
direct measure of the strength of the quasiparticle interactions or 
equivalently, in the context of the $1/N$ expansion, of the strength of the 
quantum fluctuations.

The $t$-$t^{\prime}$-$J$ model Hamiltonian can be expressed in terms of the
Hubbard operators $\chi^{ab}=|a\rangle\langle b|$ as
\begin{equation}
H=-\sum_{i,j}t_{ij}\chi^{0\mu}_{i}\chi^{\mu 0}_{j}
+{\textstyle\frac{1}{2}}J\sum_{\langle i,j\rangle}
(\chi^{\mu\nu}_{i}\chi^{\nu\mu}_{j}-\chi^{\mu\mu}_{i}\chi^{\nu\nu}_{j}) \;,
\label{eq:hamilt1}
\end{equation}
where the index 0 corresponds to a hole, the Greek indices $\mu,\nu,\ldots$ 
assume two distinct values, for a spin-up and a spin-down electron, and the 
summation convention is invoked. Here $J$ is the antiferromagnetic 
spin-exchange interaction between nearest-neighbor sites ${\langle i,j\rangle}$
on a square lattice, while for the hopping matrix elements $t_{ij}$ we assume
\begin{equation}
t_{ij}=\left\{\begin{array}{cl}
       t & \mbox{if $i,j$ are nearest neighbors,} \\
       -t^{\prime} & \mbox{if $i,j$ are next nearest neighbors,} \\
       0 & \mbox{otherwise.}
       \end{array}\right.
\label{eq:hopping}
\end{equation}
Our conventions in (\ref{eq:hopping}) incorporate opposite signs for $t$ and 
$t^{\prime}$ as dictated by quantum-chemistry calculations 
\cite{Hybertsen90,Sawatzky90} for Cu-O clusters and fits of the shape of the 
Fermi surface observed by angle-resolved photoemission spectroscopy \cite{Yu91}.
In Ref.~\onlinecite{Psaltakis93} we generalized the local constraint associated
with (\ref{eq:hamilt1}) to $\chi^{00}_{i}+\chi^{\mu\mu}_{i}=N$, where $N$ is an
arbitrary integer, and simplified the commutation properties of the 
$\chi^{ab}$'s to be those of the generators of the U(3) algebra. A generalized 
Holstein-Primakoff realization of this algebra in terms of Bose operators can 
then be employed to develop a perturbation theory based on the $1/N$ expansion, 
restoring the relevant physical value $N=1$ at the end of the calculation. For 
an average electronic density $n_{e}$ close to half-filling ($n_{e}\lesssim 1$)
and a sufficiently large $t^{\prime}$, the ground state of (\ref{eq:hamilt1}) 
in the large-$N$ limit is an unconventional metallic antiferromagnetic (AF) 
state characterized by the usual $(\pi,\pi)$ modulation of the spin 
configuration and an unusual $(\pi,-\pi)$ phase modulation of the 
condensate \cite{Psaltakis93}. It is the optical properties of this 
phase-modulated AF state that the present work is concerned with. 

The remainder of this paper is organized as follows. Section~\ref{sec:Kubo} 
contains a brief summary of the relevant Kubo formalism for the optical 
conductivity. In Sec.~\ref{sec:1/N} we derive the analytic expressions of the 
$1/N$ expansion for the Drude weight and the total optical weight, including 
leading quantum-fluctuation effects. In Sec.~\ref{sec:results} we present 
explicit numerical results for the latter quantities along with some further 
comments on the optical absorption and a discussion of the zero-sound velocity 
which is related to the Drude weight and the inverse compressibility. Our 
conclusions are summarized in Sec.~\ref{sec:concl}.

\section{Kubo formalism}
\label{sec:Kubo}

Using standard linear-response theory, the Kubo formula for the real part of 
the frequency-dependent optical conductivity $\tilde{\sigma}(\omega)$ at zero 
temperature is expressed as the sum of two physically distinct terms 
\cite{Shastry90,Scalapino92-93}
\begin{equation}
{\rm Re}[\tilde{\sigma}(\omega)]=\pi e^{2}D\delta(\omega)+\sigma(\omega) \;.
\label{eq:conductivity}
\end{equation}
The first term in (\ref{eq:conductivity}), involving the delta function at zero
frequency ($\omega=0$), is due to the free acceleration of the charge-carrying 
mass by the electric field and therefore the associated spectral weight, i.e., 
the Drude weight $D$, should vanish in the insulating state and be finite in 
the metallic state \cite{Kohn64}. $D$ measures the ratio of the density of the 
mobile charge carriers to their mass. The second term in 
(\ref{eq:conductivity}), called the optical absorption $\sigma(\omega)$, is due
to finite-frequency ($\omega>0$) optical transitions to excited quasiparticle 
states. More explicitly we have that
\begin{equation}
\sigma(\omega)=\frac{\pi}{\omega\Lambda}\sum_{m\neq 0}
|\langle m|{\rm J}|0\rangle|^{2}\delta[(E_{m}-E_{0})-\omega] \;,
\label{eq:Kubo1}
\end{equation}
where the summation is taken over a complete set of energy eigenstates
$|m\rangle$ with excitation energies $(E_{m}-E_{0})$ above the ground 
state $|0\rangle$, $\Lambda$ is the total number of lattice sites, and 
${\rm J}$ is one of the Cartesian components of the current operator ${\bf J}$ 
associated with (\ref{eq:hamilt1}),
\begin{equation}
{\bf J}=ie\sum_{i,j}t_{ij}({\bf R}_{i}-{\bf R}_{j})
\chi^{0\mu}_{i}\chi^{\mu 0}_{j} \;,
\label{eq:current}
\end{equation}
where ${\bf R}_{i}$ is the position vector for site $i$. Similarly, the Drude 
weight $D$ is given by
\begin{equation}
D=K-\frac{2}{e^{2}\Lambda}\sum_{m\neq 0}
\frac{|\langle m|{\rm J}|0\rangle|^{2}}{E_{m}-E_{0}} \;,
\label{eq:drude-weight}
\end{equation}
where $K$ is the expectation value
\begin{equation}
K=\frac{2}{z\Lambda}\langle 0|-T|0\rangle \;,
\label{eq:total-weight}
\end{equation}
with the operator $T$ defined as
\begin{equation}
T=-\sum_{i,j}t_{ij}|{\bf R}_{i}-{\bf R}_{j}|^{2}
\chi^{0\mu}_{i}\chi^{\mu 0}_{j} \;,
\label{eq:stress-tensor}
\end{equation}
$z=4$ being the coordination number of the square lattice. Integrating now over 
frequency both sides of (\ref{eq:conductivity}), using the identity 
$\int_{0}^{\infty}d\omega\,\delta(\omega)=\frac{1}{2}$ and the explicit forms 
(\ref{eq:Kubo1}) and (\ref{eq:drude-weight}), we arrive at the well-known 
optical conductivity sum rule \cite{sum-rule} 
\begin{equation}
\int_{0}^{\infty}d\omega\,{\rm Re}[\tilde{\sigma}(\omega)]
=\frac{\pi e^{2}}{2}K \;,
\label{eq:sum-rule}
\end{equation}
which, indeed, identifies $K$ as the total optical spectral weight. 

It should be noted that in lattice models involving only nearest-neighbor 
hopping, $T$ defined in (\ref{eq:stress-tensor}) coincides with the kinetic 
energy operator given by the first term on the right-hand-side of 
(\ref{eq:hamilt1}). However, the presence of an additional 
next-nearest-neighbor hopping, i.e., a hopping $t^{\prime}$ along the diagonal 
of the square unit cell as in the $t$-$t^{\prime}$-$J$ model under study, 
invalidates the latter simple result noting that: 
$|{\bf R}_{i}-{\bf R}_{j}|^{2}=2$, for next-nearest-neighbor sites $i,j$. 
Therefore in this more general case, the total optical weight $K$ given by 
(\ref{eq:total-weight}) is {\em not} just the kinetic-energy expectation value 
in the ground state. Nevertheless, by an elementary application of the 
Hellmann-Feynman theorem $K$ can still be extracted from the ground-state 
energy $\langle 0|H|0\rangle$ by differentiation with respect to the hoppings 
\cite{Baeriswyl86} $t$ and $t^{\prime}$:
\begin{equation}
K=-\frac{2}{z\Lambda}\left(t\frac{\partial}{\partial t}
+2t^{\prime}\frac{\partial}{\partial t^{\prime}}\right)
\langle 0|H|0\rangle \;.
\label{eq:Hellmann-Feynman}
\end{equation}
The identity (\ref{eq:Hellmann-Feynman}) is actually implemented in the 
following section to derive the leading terms of the $1/N$ expansion of $K$ 
from the corresponding terms of $\langle 0|H|0\rangle$ which our theory readily
provides.

\section{$\protect\bbox{1/N}$ expansion}
\label{sec:1/N}

The $1/N$ expansion of the Hamiltonian (\ref{eq:hamilt1}) and the optical 
absorption (\ref{eq:Kubo1}) around the phase-modulated AF configuration, up to 
and including terms of order $N$, has been carried out in our earlier works 
\cite{Psaltakis93,Psaltakis95}. For ease of reference, we quote in 
(\ref{eq:hamilt2})--(\ref{eq:Kubo2}) the main results necessary for our present
development. Specifically, the expansion of the Hamiltonian (\ref{eq:hamilt1}) 
reads \cite{Psaltakis93}
\begin{eqnarray}
H=&& N^{2}\Lambda E_{0}+N\Lambda E_{1} \nonumber \\
&&+N\sum_{\bf q}\left[\omega_{1}({\bf q})A^{\ast}_{\bf q}A_{\bf q}
+\omega_{2}({\bf q})B^{\ast}_{\bf q}B_{\bf q}\right] \;,
\label{eq:hamilt2}
\end{eqnarray}
where $E_{0}$ and $E_{1}$ are the classical (large-$N$ limit) and zero-point 
energy per lattice site, respectively, for the relevant physical value $N=1$:
\begin{eqnarray}
E_{0} & = & -zt^{\prime}n_{e}(1-n_{e})-\frac{zJ}{4}n_{e}^{2} \;,
\nonumber \\
& & \label{eq:E0&E1} \\
E_{1} & = & -\frac{zt^{\prime}}{2}(2-n_{e})-\frac{zJ}{4}n_{e}
+\frac{1}{2\Lambda}\sum_{\bf q}\left[
\omega_{1}({\bf q})+\omega_{2}({\bf q})\right] \;.
\nonumber
\end{eqnarray}
In the above $A_{\bf q}$ and $B_{\bf q}$ are the normal-mode operators that 
diagonalize the leading (quadratic) quantum fluctuations while 
$\omega_{n}({\bf q})$, $n=1,2$, are the dispersions of the two branches of the 
spectrum of elementary excitations. An immediate consequence of 
(\ref{eq:hamilt2}) is the following result for the $1/N$ expansion of the 
ground-state energy, up to and including terms of order $N$,
\begin{equation}
\langle 0|H|0\rangle=N^{2}\Lambda E_{0}+N\Lambda E_{1} \;.
\label{eq:energy}
\end{equation}
The corresponding expansion of the optical absorption (\ref{eq:Kubo1}) reads 
\cite{Psaltakis95} 
\begin{equation}
\sigma(\omega)=N\frac{2\pi(et)^{2}}{\omega\Lambda}\sum_{\bf q}u^{2}({\bf q})
\delta[\omega_{1}({\bf q})+\omega_{2}({\bf q})-\omega] \;.
\label{eq:Kubo2}
\end{equation}
The analytic expressions for the dimensionless function $u^{2}({\bf q})$, 
corresponding to the current matrix elements, and the dispersions 
$\omega_{n}({\bf q})$, $n=1,2$, are rather involved and are summarized in the 
Appendix. The result (\ref{eq:Kubo2}) has been extensively discussed in our 
earlier work of Ref.~\onlinecite{Psaltakis95} while some further comments are 
added in Sec.~\ref{sec:results}. 

At present, an important point to note in (\ref{eq:Kubo2}) is the lack of a 
classical contribution, i.e., a term of order $N^{2}$, to $\sigma(\omega)$. 
In other words, the finite-frequency component $\sigma(\omega)$ of the optical 
conductivity is due {\em exclusively} to the quantum fluctuations whose leading
contribution, within the present $1/N$ expansion scheme, is of order $N$. This 
conclusion is consistent with similar observations made Bang and Kotliar 
\cite{Bang93} for the finite-frequency optical conductivity of the simple 
$t$-$J$ model treated by a slave-boson diagrammatic $1/N$ expansion technique.

The derivation of the $1/N$ expansion of the Drude weight $D$ and the total 
optical weight $K$, up to and including terms of order $N$, proceeds now in two
steps: (i) The result (\ref{eq:energy}) is used to implement identity 
(\ref{eq:Hellmann-Feynman}). (ii) The second term on the right-hand-side of 
(\ref{eq:drude-weight}) is expressed initially as the integral of 
(\ref{eq:Kubo1}) over positive frequencies and then (\ref{eq:Kubo2}) is 
exploited to obtain immediately its $1/N$ expansion. Taking into account 
(\ref{eq:E0&E1}), the final analytic results of this straightforward procedure 
may be written in the form
\begin{eqnarray}
D & = & N^{2}D_{0}+ND_{1} \;, 
\label{eq:D-expansion} \\
K & = & N^{2}K_{0}+NK_{1} \;,
\label{eq:K-expansion} 
\end{eqnarray}
where $D_{0}$ and $K_{0}$ are the classical (large-$N$ limit) contributions to
the Drude weight and the total optical weight, respectively,
\begin{equation}
D_{0}=K_{0}=-\frac{2}{z}\left(t\frac{\partial}{\partial t}
+2t^{\prime}\frac{\partial}{\partial t^{\prime}}\right)E_{0}
=4t^{\prime}n_{e}(1-n_{e}) \;,
\label{eq:D0&K0}
\end{equation}
while $D_{1}$ and $K_{1}$ are the corresponding contributions due to leading
quantum-fluctuation effects, 
\begin{eqnarray}
K_{1} & = & -\frac{2}{z}\left(t\frac{\partial}{\partial t}
+2t^{\prime}\frac{\partial}{\partial t^{\prime}}\right)E_{1} \;, 
\label{eq:K1} \\
D_{1} & = & K_{1}-\frac{4t^{2}}{\Lambda}\sum_{\bf q}\frac{u^{2}({\bf q})}
{\omega_{1}({\bf q})+\omega_{2}({\bf q})} \;.
\label{eq:D1}
\end{eqnarray}
The equality $D_{0}=K_{0}$ quoted in (\ref{eq:D0&K0}) is a consequence of the 
absence of quantum fluctuations in the large-$N$ limit and implies that in this
classical approximation, described by the terms of order $N^{2}$, the 
$\delta$-function Drude peak centered at zero frequency ($\omega=0$) carries 
the total optical weight. However, as noted earlier on, the leading quantum 
fluctuations, described by the terms of order $N$, already give rise to optical
absorption and hence distribute part of the total optical weight to 
finite frequencies ($\omega>0$). The presence of the terms of order $N$ in 
(\ref{eq:D-expansion}) and (\ref{eq:K-expansion}) leads therefore to $D<K$; a 
generic property anticipated also from (\ref{eq:drude-weight}) and 
(\ref{eq:D1}).

In view of (\ref{eq:E0&E1}) and the explicit forms quoted in the Appendix, the 
derivatives with respect to $t$ and $t^{\prime}$ appearing in (\ref{eq:K1}) can
be carried out analytically. The remaining wave vector integrations over the 
square Brillouin zone in (\ref{eq:K1}) and (\ref{eq:D1}), as well as in 
(\ref{eq:Kubo2}), can then be performed numerically. Explicit numerical results
derived from (\ref{eq:D-expansion})--(\ref{eq:D1}) in the way just described 
are presented in the following section.

\section{Explicit results}
\label{sec:results}

For completeness and better insight to the results for the Drude weight and the
total optical weight that will be presented shortly, it is instructive to begin
our discussion here with some examples of the optical absorption line shape 
$\sigma(\omega)$. In Fig.~\ref{fig:absorption} we draw $\sigma(\omega)$, 
determined from (\ref{eq:Kubo2}) with $N=1$, for typical values of the 
dimensionless ratios $\varepsilon=t^{\prime}/t$, $t/J$ and the hole 
concentration $(1-n_{e})$ that are relevant for the copper-oxide layers 
\cite{Hybertsen90,Sawatzky90,Yu91}. As noted in Ref.~\onlinecite{Psaltakis95}, 
the limiting value $\sigma(\omega\rightarrow 0)$ is finite, at finite hole 
doping, while for small doping values (such as the 10\% hole doping considered 
in Fig.~\ref{fig:absorption}) the high-frequency divergence of $\sigma(\omega)$
comes from the ${\bf q}=(\pi/2,\pi/2)$ point of the Brillouin zone. The 
frequency position of this peak (divergence) is then determined as
\begin{equation}
[\omega_{1}({\bf q})+\omega_{2}({\bf q})]_{{\bf q}=(\pi/2,\pi/2)}=
zt^{\prime}(2-n_{e})+\frac{zJ}{2}n_{e} \;,
\label{eq:van Hove}
\end{equation}
and apparently is independent of the nearest-neighbor hopping $t$. For the 
parameter values corresponding to the solid line of 
Fig.~\ref{fig:absorption}(a), the peak position of $\sigma(\omega)$ 
calculated from (\ref{eq:van Hove}) is given by $\omega=3.78 J$. The latter 
value, with an estimated \cite{Aeppli89,Rossat91,Shamoto93} 
$J\approx 0.13$--$0.16\,\mbox{eV}$, accounts aptly for the 
$\omega\approx 0.5\,\mbox{eV}$ peak of the midinfrared band observed in optical
conductivity measurements \cite{Cooper90,Orenstein90,Uchida91} in doped 
antiferromagnets. 

From the explicit form (\ref{eq:van Hove}) follows that close to half-filling 
and for parameters satisfying $2t\varepsilon/J<1$ the peak of $\sigma(\omega)$ 
shifts slowly to lower frequencies upon hole doping, i.e., with increasing 
$(1-n_{e})$. Such a systematic shift of the peak of the midinfrared band is, 
indeed, observed in the afore-mentioned optical conductivity measurements. 
Interesting enough, close to half-filling the same condition 
$2t\varepsilon/J<1$ has been shown in Ref.~\onlinecite{Psaltakis93} to ensure 
the softening of the magnon velocity upon hole doping; a trend that is also 
observed in inelastic neutron-scattering measurements from doped 
antiferromagnets \cite{Aeppli89,Rossat91}. The present theory provides 
therefore a context for the qualitative understanding of both experimental 
observations. Concluding our comments on $\sigma(\omega)$ we note that 
(\ref{eq:conductivity}) and (\ref{eq:sum-rule}) yield 
$\int_{0}^{\infty}d\omega\,\sigma(\omega)=\frac{\pi e^{2}}{2}(K-D)$. Therefore,
the spectral weight carried by the finite-frequency component $\sigma(\omega)$ 
is given by the difference $(K-D)$ and hence its dependence on the parameters 
$\varepsilon$, $t/J$, and $n_{e}$ follows from that of $D$ and $K$.

Let us now consider the large-$N$ limit contributions $D_{0}$ and $K_{0}$ to 
the Drude weight and the total optical weight, respectively. The equality of 
these two quantities has already been discussed following (\ref{eq:D1}). Here we
note that the vanishing overlap between the opposite sublattice spin states
in the AF configuration, along with the absence of quantum fluctuations in the 
large-$N$ limit, leaves the direct hopping $t^{\prime}$ between same sublattice
sites as the only relevant process of charge transport in this classical 
approximation. This argument makes plausible the independence of 
$D_{0}$, $K_{0}$ from $t$ and $J$ seen in (\ref{eq:D0&K0}).

Furthermore, the $n_{e}$-dependence in (\ref{eq:D0&K0}) implies that: 
(i) $D_{0}=K_{0}\propto n_{e}$, for $n_{e}\rightarrow 0$, and 
(ii) $D_{0}=K_{0}\propto (1-n_{e})$, for $n_{e}\rightarrow 1$.
The linear increase of $D_{0}$ and $K_{0}$ with small electron density $n_{e}$ 
away from the empty-lattice limit ($n_{e}=0$) is, of course, an expected 
behavior in this ``free'' electron gas regime. On the other hand, the linear 
increase of $D_{0}$ and $K_{0}$ with small hole concentration $(1-n_{e})$ away 
from the half-filled-band limit ($n_{e}=1$) is an important consequence of the 
local constraint which prohibits the occupancy of any lattice site by more than 
one electron and leads inevitably to the Mott insulator at half-filling 
($D_{0}=0=K_{0}$, for $n_{e}=1$.) The latter property, typified in point-(ii)
above, serves to interpret the ``free'' charge carriers near half-filling as 
being holes rather than electrons. Then, the position of the maximum of $D_{0}$ 
located at quarter-filling ($n_{e}=\frac{1}{2}$) provides an estimate of the 
critical amount of doping at which the character of the charge carriers changes
from holelike to electronlike with increasing $(1-n_{e})$. The present 
large-$N$ limit results capture already generic features of the 
$n_{e}$-dependence of the Drude weight and the total optical weight that are 
derived by exact diagonalization of the simple $t$-$J$ model and the large-$U$ 
Hubbard model on small clusters \cite{Stephan92,Dagotto92,review94}. The 
reason can be traced to the fact that our generalized Holstein-Primakoff 
realization \cite{Psaltakis93} resolves explicitly the local constraint and 
thus the ensuing $1/N$ expansion incorporates already part of the strong 
correlation effects, implied by this constraint, in the leading order.

As a check of consistency of the quantum-liquid description emerging from the 
present $1/N$ expansion scheme it is worthwhile at this point to discuss 
briefly the zero-sound velocity, i.e., the velocity of the long-wavelength 
charge excitations, in the physically relevant regime near half-filling 
($n_{e}\rightarrow 1$). As demonstrated in detail in 
Ref.~\onlinecite{Psaltakis93}, a separation of spin and charge degrees of 
freedom sets in progressively and becomes {\em asymptotically} exact as the 
half-filled-band limit is approached. The latter limit permits then an accurate
classification of the elementary excitations in terms of a mode that describes 
the spin excitations and a mode that describes the charge excitations. At 
long-wavelengths ($|{\bf q}|\rightarrow 0$), in particular, the latter mode 
corresponds to the lower-energy branch of the twofold excitation spectrum 
given by $\omega_{2}({\bf q})=c_{2}|{\bf q}|$. Thus near half-filling 
($n_{e}\rightarrow 1$) the velocity of the long-wavelength charge excitations 
is given by $c_{2}$. On the other hand, in the context of conventional 
quantum-liquid theory \cite{Pines89} the zero-sound velocity may be expressed 
in terms of the Drude weight $D_{0}$ and the inverse compressibility 
$1/\kappa=n_{e}^{2}E_{0}^{\prime\prime}\,$, where 
$E_{0}^{\prime\prime}=\partial^{2}E_{0}/\partial n_{e}^{2}\,$, as 
$\sqrt{D_{0}/(n_{e}^{2}\kappa)}=\sqrt{D_{0}E_{0}^{\prime\prime}}\,$. Hence the 
consistency of the quantum-liquid description requires the validity of the 
identity: $c_{2}=\sqrt{D_{0}E_{0}^{\prime\prime}}\,$, for $n_{e}\rightarrow 1$.
The asymptotic form of $c_{2}$ for $n_{e}\rightarrow 1$ has been derived in 
Ref.~\onlinecite{Psaltakis93} by considering explicitly the long-wavelength 
limit of $\omega_{2}({\bf q})$. Here, using (\ref{eq:E0&E1}) and 
(\ref{eq:D0&K0}) to calculate $\sqrt{D_{0}E_{0}^{\prime\prime}}\,$ and 
comparing with the latter result for $c_{2}$ one can easily verify that indeed
\cite{Schulz90}
\begin{eqnarray}
c_{2}=\sqrt{D_{0}E_{0}^{\prime\prime}}=2\varepsilon
\biggl[2ztJ\biggl(\frac{t}{J}-\frac{1}{4\varepsilon}\biggr)
(&&1-n_{e})\biggr]^{1/2} \nonumber \\
&& \nonumber \\
&& \mbox{for}\;\; n_{e}\rightarrow 1 \;.
\label{eq:zero-sound} 
\end{eqnarray}
In view of the radically different way in which $c_{2}$ and 
$\sqrt{D_{0}E_{0}^{\prime\prime}}\,$ are derived, the relation 
(\ref{eq:zero-sound}) provides a stringent consistency check of our theory and 
serves to identify unambiguously $c_{2}$ as the slow zero-sound velocity for 
$n_{e}\rightarrow 1$. Furthermore, from (\ref{eq:E0&E1}) follows that 
$E_{0}^{\prime\prime}\,$ is independent of $n_{e}$. Hence (\ref{eq:zero-sound})
reveals that the physical reason for the softening of the velocity $c_{2}$ of 
the lower-energy branch $\omega_{2}$ near the Mott metal-insulator transition 
at half-filling ($c_{2}\propto\sqrt{1-n_{e}}\rightarrow 0$, for 
$n_{e}\rightarrow 1$) is the collapse of the Drude weight which in turn is a 
direct consequence of the local constraint. It should be noted that this slow 
zero-sound mode near the Mott transition appears also in slave-boson treatments
of the local constraint when the fluctuations of the associated statistical 
gauge field are taken into account beyond the mean-field approximation 
\cite{Rodriguez94}. Concluding our discussion of the zero-sound velocity we note
that away from the asymptotic regime $n_{e}\rightarrow 1$ and with increasing 
hole concentration, $\omega_{2}$ (as well as $\omega_{1}$) involves an 
increasingly strong hybridization between spin and charge degrees of freedom 
and consequently $c_{2}$ corresponds no longer to the velocity of the 
long-wavelength charge excitations.

Having analyzed the classical (large-$N$ limit) contributions to the Drude 
weight and the total optical weight, we consider in the remainder of this 
section the complete results for $D$ and $K$ that include the leading 
quantum-fluctuation corrections, as determined from (\ref{eq:D-expansion}) and 
(\ref{eq:K-expansion}), respectively, with $N=1$. In Fig.~\ref{fig:D&K-0.45} 
(for $\varepsilon=0.45$) and Fig.~\ref{fig:D&K-0.40} (for $\varepsilon=0.40$) 
we draw as a function of the hole concentration $(1-n_{e})$: (a) the Drude 
weight $D$, (b) the total optical weight $K$, and (c) the fraction $D/K$, for 
$t/J=1.0$ (solid line) or $t/J=1.5$ (dashed line). For comparison, in each 
figure we also depict by dotted line the classical (large-$N$ limit) result for
the drawn quantity: (a) $D_{0}$, (b) $K_{0}$, and (c) $D_{0}/K_{0}$. We remind 
that the latter classical results, determined from (\ref{eq:D0&K0}), are 
independent of the ratio $t/J$. Our main observations from 
Fig.~\ref{fig:D&K-0.45} and Fig.~\ref{fig:D&K-0.40} are summarized as follows.

(i) $D$ and $K$ increase linearly with small hole concentration $(1-n_{e})$ 
away from the Mott metal-insulator transition point at half-filling 
($n_{e}=1$), consistent with the notion that the ``free'' charge carriers in 
this regime are the holes. However, as a result of the leading quantum 
fluctuations, $D$ and $K$ are no longer equal. Close to half-filling and for 
the typical values of $\varepsilon$ and $t/J$ considered, the fraction $D/K$ is
reduced from its ``classical'' value of 1 to about 0.5 and is almost 
independent of $(1-n_{e})$ in the doping range of interest; see the solid and 
dashed lines in Fig.~\ref{fig:D&K-0.45}(c) and Fig.~\ref{fig:D&K-0.40}(c). 
These results are consistent with optical conductivity measurements 
\cite{Cooper90,Orenstein90,Uchida91} in doped antiferromagnets. In particular,
for $\varepsilon=0.45$, $t/J=1.0$, corresponding to the solid line of 
Fig.~\ref{fig:D&K-0.45}(c), and a typical 10\% hole doping we have: $D/K=0.52$,
corresponding to a mass enhancement factor $K/D=1.92$. The latter value 
compares reasonably well with the experimental estimate \cite{Orenstein90} 
$(K/D)_{\rm exp}\approx 2.3$. Furthermore, our typical value $D/K=0.52$ is
compatible with the corresponding prediction of 0.6 derived by exact 
diagonalization \cite{Dagotto92} and anyon techniques \cite{Tikofsky92} for the
simple $t$-$J$ model, and implies that 52\% of the total optical weight resides
at the zero-frequency Drude peak $\pi e^{2}D\delta(\omega)$ while the rest is 
carried by the finite-frequency component $\sigma(\omega)$. It should be noted 
that the strength of the quasiparticle interactions, as inferred from the mass 
enhancement factor $K/D\approx 2.0$, is not exceptionally large 
\cite{Orenstein90}. This fact, however, is not surprising noting that important
correlations effects, implied by the local constraint, are already absorbed 
into the charge carrier density which near half-filling is given by the hole 
concentration $(1-n_{e})$ that vanishes at $n_{e}=1$, leading to the Mott 
metal-insulator transition.

(ii) At higher doping values, $D$ and $K$ eventually saturate and then start to
decrease. However, the leading quantum fluctuations shift the maximum of $D$ 
and $K$ away from its ``classical'' location at quarter-filling towards 
half-filling. This shift is small and slightly different for $D$ and $K$. As 
noted earlier on and emphasized by Dagotto {\it et al.} \cite{Dagotto92}, the 
position of the maximum of $D$ provides an estimate of the doping value at 
which the charge carriers change from holelike to electronlike with increasing 
$(1-n_{e})$, hence at roughly the same position the Hall coefficient $R_{H}$ 
should change sign from positive to negative. For $\varepsilon=0.45$ (0.40) and 
$t/J=1.0$, corresponding to the solid line of Fig.~\ref{fig:D&K-0.45}(a) 
[Fig.~\ref{fig:D&K-0.40}(a)], the maximum of $D$ occurs at the hole 
concentration $(1-n_{e})=0.44$ (0.36). The latter result is then compatible 
with numerical and analytical calculations \cite{Hall-theory} of $R_{H}$, in 
the context of the simple $t$-$J$ model, that predict a sign change at about 
0.40--0.33 hole concentration, and with Hall effect measurements 
\cite{Hall-experi} in doped antiferromagnets that report a sign change of 
$R_{H}$ at about 0.3 hole concentration. 

(iii) We cannot approach the ``free'' electron gas regime close to the 
empty-lattice limit ($n_{e}=0$) because the phase-modulated AF configuration, 
around which the present $1/N$ expansion is carried out, becomes unstable 
beyond a critical value of the hole concentration $(1-n_{e})$. The intervening 
instability is reflected in the elementary excitations whereby the velocity 
$c_{2}$ of the lower-energy branch $\omega_{2}$ becomes zero at the critical 
doping value \cite{Psaltakis93}. This results in a very rapid decrease of $D$, 
$K$, and $D/K$ in the immediate neighborhood of the latter doping value; a 
behavior clearly seen in Fig.~\ref{fig:D&K-0.45} and Fig.~\ref{fig:D&K-0.40}. 
In this critical doping regime, the magnitude of the (negative) leading 
quantum-fluctuation corrections $D_{1}$ and $K_{1}$ becomes even larger than 
that of the classical (large-$N$ limit) terms $D_{0}$ and $K_{0}$ and therefore
the $1/N$ expansion breaks down.

(iv) Finally, by comparing the solid and dashed lines in 
Fig.~\ref{fig:D&K-0.45} and Fig.~\ref{fig:D&K-0.40} we conclude that at any 
given doping value $D$, $K$, and $D/K$ all increase with increasing 
$t^{\prime}$ and/or $J$. This trend is consistent with similar results derived 
by exact diagonalization of the $t$-$J$ model, extended to include the 
so-called three-site term \cite{Stephan92}, and of the Hubbard model 
\cite{Dagotto92}. Therefore, this trend should be regarded as being generic for 
models involving a direct or an effective hopping between same sublattice sites.

\section{Conclusions}
\label{sec:concl}

In this paper we have presented a detailed study of the  Drude weight $D$, the 
total optical weight $K$, and the fraction $D/K$ that defines the inverse of 
the mass enhancement factor, in the phase-modulated AF state of the 
$t$-$t^{\prime}$-$J$ model (\ref{eq:hamilt1})--(\ref{eq:hopping}). The generic 
features of these quantities observed in optical conductivity measurements 
\cite{Cooper90,Orenstein90,Uchida91} in doped antiferromagnets are 
qualitatively reproduced when the leading quantum-fluctuation effects, around 
the afore-mentioned semiclassical ground state, are taken into account within a
suitable $1/N$ expansion. 

Specifically, $D$ and $K$ increase linearly with small hole doping away from 
the Mott metal-insulator transition at half-filling, consistent with the notion
that the ``free'' charge carriers in this regime are the holes. Our theoretical
prediction of a mass enhancement factor $K/D\approx 2.0$ that is almost 
independent of the hole concentration, in the doping range of interest, 
compares reasonably well with the corresponding experimental estimate. 
\cite{Orenstein90} With increasing hole doping $D$ (and $K$) reaches eventually
a maximum at a value that is compatible with measurements \cite{Hall-experi} of
the critical doping at which the Hall coefficient $R_{H}$ changes sign from 
positive to negative, signaling a change in the character of the charge 
carriers from holelike to electronlike. Furthermore, for typical parameter 
values the peak position (\ref{eq:van Hove}) of the finite-frequency component 
$\sigma(\omega)$ of the optical conductivity accounts aptly for the 
experimentally observed $0.5\,\mbox{eV}$ peak of the midinfrared band 
\cite{Cooper90,Orenstein90,Uchida91}.

Finally, the slow zero-sound velocity near the Mott metal-insulator transition 
point at half-filling is shown to identify with the velocity of the 
lower-energy branch of the twofold excitation spectrum, thus providing a 
stringent consistency check of the quantum-liquid description emerging from the
present $1/N$ expansion scheme.

\section*{Acknowledgments}

I would like to thank N. Papanicolaou for valuable discussions. This work was 
supported by Grant No. CHRX-CT93-0332 from the EEC and Grant No. 91E$\Delta$631
from the Greek Secretariat for Research and Technology.

\appendix
\section*{}

In this Appendix we summarize the analytic expressions for the dispersions 
$\omega_{n}({\bf q})$, $n=1,2$, of the two branches of the spectrum of 
elementary excitations, and the dimensionless function $u^{2}({\bf q})$, 
corresponding to the current matrix elements, that enter the optical absorption
(\ref{eq:Kubo2}) and the leading quantum-fluctuation corrections (\ref{eq:K1}),
(\ref{eq:D1}) for the Drude weight and the total optical weight.

Following the notation conventions of Ref.~\onlinecite{Psaltakis93}, the 
dispersions $\omega_{n}({\bf q})$, $n=1,2$, are given by
\begin{eqnarray}
\omega_{1}^{2}({\bf q}) & = & R({\bf q})+2\sqrt{S({\bf q})} \;, 
\nonumber \\
& & \label{eq:omega} \\
\omega_{2}^{2}({\bf q}) & = & R({\bf q})-2\sqrt{S({\bf q})} \;,
\nonumber
\end{eqnarray}
with
\begin{equation}
R({\bf q})=\left[\omega^{(+)}_{\bf q}\right]^{2}
+\left[\omega^{(-)}_{\bf q}\right]^{2}
+\tau_{\bf q}^{2}
-\lambda_{\bf q}^{2} 
-\left[\nu^{(+)}_{\bf q}\right]^{2}
-\left[\nu^{(-)}_{\bf q}\right]^{2},
\label{eq:R(q)}
\end{equation}
and 
\begin{eqnarray}
S({\bf q})=&&\left[\omega^{(+)}_{\bf q}\omega^{(-)}_{\bf q}
-\lambda_{\bf q}\nu^{(-)}_{\bf q}\right]^{2}
+\left[\omega^{(+)}_{\bf q}\tau_{\bf q}
-\nu^{(+)}_{\bf q}\lambda_{\bf q}\right]^{2} \nonumber \\
&&-\left[\omega^{(-)}_{\bf q}\nu^{(+)}_{\bf q}
-\nu^{(-)}_{\bf q}\tau_{\bf q}\right]^{2} \;,
\label{eq:S(q)-AF}
\end{eqnarray}
where for an arbitrary wave vector ${\bf q}=(q_{x},q_{y})$ the explicit forms of
the coefficients $\omega^{(\pm)}_{\bf q}$, $\tau_{\bf q}$, $\lambda_{\bf q}$, 
and $\nu^{(\pm)}_{\bf q}$ read
\begin{eqnarray}
&&\lambda_{\bf q}=\frac{zt^{\prime}}{4}\frac{n_{e}^{2}}{(1-n_{e})}
(1-\cos q_{x}\cos q_{y}) \nonumber \\
&&\qquad\;\, +\frac{zt^{\prime}}{2}n_{e}(1+\cos q_{x}\cos q_{y}) \;, \\
\nonumber \\
&&\tau_{\bf q}=\lambda_{\bf q}-\frac{zJ}{4}n_{e}
\left[1+\frac{1}{2}(\cos q_{x}+\cos q_{y})\right] \;, \\
\nonumber \\
&&\nu^{(+)}_{\bf q}=\lambda_{\bf q}
-\frac{zJ}{4}n_{e}(\cos q_{x}+\cos q_{y}) \;, \\
\nonumber \\
&&\nu^{(-)}_{\bf q}=\frac{zt}{4}n_{e}(\cos q_{x}-\cos q_{y}) \;, \\
\nonumber \\
&&\omega^{(+)}_{\bf q}=\nu^{(+)}_{\bf q}+\lambda_{\bf q}-\tau_{\bf q}
+zt^{\prime}(1-n_{e})(1-\cos q_{x}\cos q_{y}) \;, \nonumber \\ 
&& \\
\nonumber \\
&&\omega^{(-)}_{\bf q}=\nu^{(-)}_{\bf q}
-\frac{zt}{2}(1-n_{e})(\cos q_{x}-\cos q_{y}) \;.
\end{eqnarray}

The analytic expression for the function $u^{2}({\bf q})$ is very involved. 
However, the result can be presented in a compact manner with the help of a 
four-component matrix notation \cite{Psaltakis93}. To this end we introduce the
$4\times 4$ matrices $\rho_{1}$, $\rho_{2}$, and $\rho_{3}$, defined as 
$\rho_{i}=\sigma_{i}\otimes 1$, $i=1,2,3$, with the $\sigma_{i}$ denoting the 
ordinary $2\times 2$ Pauli matrices. The same symbols $\sigma_{i}$ will also be
used in the following to denote the $4\times 4$ matrices $1\otimes\sigma_{i}$, 
$i=1,2,3$, for notational simplicity. We then have
\begin{equation}
u^{2}({\bf q})={\textstyle\frac{1}{2}}
[u_{x}^{2}({\bf q})+u_{y}^{2}({\bf q})] \;,
\label{eq:u(q)}
\end{equation}
where $u_{\alpha}^{2}({\bf q})$, $\alpha=x,y$, are positive-definite 
dimensionless functions expressed as traces over $4\times 4$ matrices. 
Specifically,
\begin{eqnarray}
u_{\alpha}^{2}({\bf q})=&& -\left(\frac{1}{4}\right)^{2}
\frac{1}{\omega_{1}({\bf q})\omega_{2}({\bf q})S({\bf q})}\bigl\{ 
\nonumber \\
&& +{\rm Tr}[v_{\alpha}({\bf q})A({\bf q},\omega_{1}({\bf q}))  
v_{\alpha}({\bf q})A({\bf q},-\omega_{2}({\bf q}))]               
\nonumber \\
&& +{\rm Tr}[v_{\alpha}({\bf q})A({\bf q},\omega_{2}({\bf q}))  
v_{\alpha}({\bf q})A({\bf q},-\omega_{1}({\bf q}))]\bigr\} \;,    
\nonumber \\
&& 
\label{eq:u_{alpha}(q)}
\end{eqnarray}
where $v_{\alpha}({\bf q})$, $\alpha=x,y$, and $A({\bf q},\omega)$ are the 
$4\times 4$ real matrices defined below:
\begin{eqnarray}
v_{x}({\bf q})=\sin q_{x}\biggl[&&
-\varepsilon\left(1-\frac{3n_{e}}{2}\right)\cos q_{y}
-\frac{1}{2}\left(1-\frac{3n_{e}}{2}\right)\sigma_{3}
\nonumber \\
&&+\frac{\varepsilon}{2}n_{e}\cos q_{y}\sigma_{1}
+\frac{i}{4}n_{e}\rho_{1}\sigma_{2}\biggr] \;, 
\label{eq:v_{x}(q)} \\
&& \nonumber \\
v_{y}({\bf q})=\sin q_{y}\biggl[&&
-\varepsilon\left(1-\frac{3n_{e}}{2}\right)\cos q_{x}
+\frac{1}{2}\left(1-\frac{3n_{e}}{2}\right)\sigma_{3}
\nonumber \\
&&+\frac{\varepsilon}{2}n_{e}\cos q_{x}\sigma_{1}
-\frac{i}{4}n_{e}\rho_{1}\sigma_{2}\biggr] \;,
\label{eq:v_{y}(q)}
\end{eqnarray}
\begin{equation}
A({\bf q},\omega)=[\omega+{\cal H}({\bf q})]
[\omega^{2}-R({\bf q})+2{\cal M}({\bf q})] \;, \;\;\, \quad
\label{eq:A(q,omega)}
\end{equation}
with
\begin{eqnarray}
{\cal H}({\bf q})=&&\omega^{(+)}_{\bf q}\rho_{3}
+\omega^{(-)}_{\bf q}\rho_{3}\sigma_{3}
+\tau_{\bf q}\rho_{3}\sigma_{1}
\nonumber \\
&&+i\lambda_{\bf q}\rho_{2}
+i\nu^{(+)}_{\bf q}\rho_{2}\sigma_{1}
+i\nu^{(-)}_{\bf q}\rho_{2}\sigma_{3} \;,
\label{eq:H(q)-AF}
\end{eqnarray}
and
\begin{eqnarray}
{\cal M}({\bf q})=&&\left[\omega^{(+)}_{\bf q}\omega^{(-)}_{\bf q}
-\lambda_{\bf q}\nu^{(-)}_{\bf q}\right]\sigma_{3}
+\left[\omega^{(+)}_{\bf q}\tau_{\bf q}
-\nu^{(+)}_{\bf q}\lambda_{\bf q}\right]\sigma_{1} \nonumber \\
&&+i\left[\omega^{(-)}_{\bf q}\nu^{(+)}_{\bf q}
-\nu^{(-)}_{\bf q}\tau_{\bf q}\right]\rho_{1}\sigma_{2} \;.
\label{eq:M(q)-AF}
\end{eqnarray}

In conclusion it is worth emphasizing that the dispersions 
$\omega_{n}({\bf q})$, $n=1,2$, defined in (\ref{eq:omega}) as well as the 
function $u^{2}({\bf q})$ defined in (\ref{eq:u(q)}), depend on ${\bf q}$ only 
through the factors $\cos q_{x}$ and $\cos q_{y}$ and are symmetric under the 
interchange of variables $(q_{x},q_{y})$: 
$\omega_{n}(q_{x},q_{y})=\omega_{n}(q_{y},q_{x})$, $n=1,2$, and 
$u^{2}(q_{x},q_{y})=u^{2}(q_{y},q_{x})$.

\begin{figure}
\caption{Optical absorption $\sigma(\omega)$ (in units of $e^{2}/\hbar$) over 
hole concentration $(1-n_{e})$ vs frequency for $1-n_{e}=0.10$, $t/J=1.0$ 
(solid line) or $t/J=1.5$ (dashed line), and (a) $\varepsilon=0.45$, (b) 
$\varepsilon=0.40$.}
\label{fig:absorption}
\end{figure}

\begin{figure}
\caption{For $\varepsilon=0.45$ and $t/J=1.0$ (solid line) or $t/J=1.5$ 
(dashed line), including leading quantum-fluctuation effects: (a) Drude weight 
$D$ vs hole concentration. (b) Total optical weight $K$ vs hole concentration. 
(c) Fraction $D/K$ vs hole concentration. In each figure, the dotted line
is the classical (large-$N$ limit) result for the drawn quantity and is 
independent of $t/J$ according to Eq.~(\protect\ref{eq:D0&K0}).}
\label{fig:D&K-0.45}
\end{figure}

\begin{figure}
\caption{For $\varepsilon=0.40$ and $t/J=1.0$ (solid line) or $t/J=1.5$ 
(dashed line), including leading quantum-fluctuation effects: (a) Drude weight 
$D$ vs hole concentration. (b) Total optical weight $K$ vs hole concentration. 
(c) Fraction $D/K$ vs hole concentration. In each figure, the dotted line
is the classical (large-$N$ limit) result for the drawn quantity and is 
independent of $t/J$ according to Eq.~(\protect\ref{eq:D0&K0}).}
\label{fig:D&K-0.40}
\end{figure}
\end{document}